\documentclass[aps,pra,reprint,showpacs,longbibliography,superscriptaddress]{revtex4-2}
\usepackage{amsmath,amssymb,amsfonts,bm}
\usepackage{graphicx}
\usepackage{hyperref}
\usepackage{physics}
\usepackage{bbold}
\usepackage{mathtools}
\usepackage{tikz}
\usepackage{subfigure} 
\hypersetup{
  colorlinks = true,
  linkcolor  = blue,
  citecolor  = blue,
  urlcolor   = blue,
}

\begin{document}

\title{Interaction-Enabled Hartree Fixed Points in Fermionic Resetting Dynamics}

\author{Jishad Kumar}
\affiliation{%
  MSP group, Department of Applied Physics, Aalto University,\\
  P.O. Box 15600, FI-00076 Aalto, Espoo, Finland
}

\author{Achilleas Lazarides}
\affiliation{%
  Interdisciplinary Centre for Mathematical Modelling and Department of Mathematical Sciences,\\
  Loughborough University, Loughborough, Leicestershire LE11 3TU, United Kingdom
}

\author{Tapio Ala-Nissila}
\affiliation{%
  MSP group, Department of Applied Physics, Aalto University,\\
  P.O. Box 15600, FI-00076 Aalto, Espoo, Finland
}
\affiliation{%
  Interdisciplinary Centre for Mathematical Modelling and Department of Mathematical Sciences,\\
  Loughborough University, Loughborough, Leicestershire LE11 3TU, United Kingdom
}

\date{\today}

\begin{abstract}
In resetting dynamics a system is repeatedly coupled to and decoupled from ancillary degrees of freedom that are reinitialized between interactions. This provides a versatile route to engineer nonequilibrium steady states and constitutes a powerful and analytically transparent framework for studying nonequilibrium dynamics in quadratic fermionic models. The baseline noninteracting resetting scheme yields an affine evolution for the subsystem single-particle density matrix (SPDM), with a clear operational interpretation: a finite environment block $E$ mediates the interaction between the subsystem $S$ and an ideal external thermal reservoir. In this work we develop a controlled extension of such a framework to weakly interacting systems.  We introduce a Hartree mean-field treatment of density–density interactions that preserves closure of the SPDM dynamics while producing genuinely nonlinear behavior.  We further construct a completely positive (CP-safe) Gaussian Lindblad embedding that reproduces the resetting
dynamics in the noninteracting limit and yields a continuous-time representation of environmental thermalization when interactions are present. Our analytical results are complemented by numerical studies of a ring segmentation geometry and a minimal two-site model, revealing interaction-enabled steady states that cannot be obtained in any purely quadratic setting. Together, these results establish a general and physically consistent route for incorporating weak interactions into resetting-based approaches to open quantum systems.
\end{abstract}

\maketitle

\section{Introduction}
\label{sec:intro}
Open quantum systems are traditionally modeled by weak coupling to macroscopic reservoirs, leading in suitable limits to Markovian master equations of Gorini--Kossakowski--Sudarshan--Lindblad (GKLS) form~\cite{lindblad1976,gorini1976,spohn1976,davies1977,breuer2002,rivas2012,chruscinski2017}. In many mesoscopic and cold-atom architectures, however, the environment is finite and structured. Repeated-interaction and resetting schemes, in which a system sequentially interacts with specific ancillary degrees of freedom that are reinitialized between interactions, have emerged as a flexible alternative description~\cite{attalpautrat2003,attal2006,grimmer2017}. Such schemes have been widely employed to study quantum thermodynamics~\cite{Strasberg2017PRX,strasberg2019,Rodrigues2019PRL}, transport~\cite{Kamiya2013JPSJ,Erbanni2023PRA}, quantum information processing~\cite{Ciccarello2022PhysRep,Cusumano2022Entropy,Cattaneo2022OSID,Campbell2021EPL,Bozkurt2025PRR}, stochastic quantum 
evolution \cite{jacquier2025} and even quantum 
computing algorithms \cite{northcote2025}.

Resetting protocols in open quantum systems have recently emerged as a conceptually simple and analytically powerful method for constructing nonequilibrium steady states in quadratic fermionic models~\cite{vieira2020}. In this noninteracting resetting framework, a finite environment block $E$ of the system-environment density matrix is periodically reinitialized to a thermal target single-particle density matrix (SPDM) $F_E$ (Fermi distribution) while the system block $S$ evolves unitarily. Further, at reset the $S +E$ blocks either evolve unitarily between resets or are set to zero (cf. Fig. 1 in Ref. \onlinecite{vieira2020}). This induces an affine linear map on the subsystem  SPDM, enabling exact characterization of relaxation, transport, and dynamical fixed points without explicit modeling of microscopic reservoir dynamics. The reset dynamics induce an effective interaction that leads to homogenized steady-states, which in some cases correspond to environmentally thermalized or pseudothermalized states.

While noninteracting quadratic resetting models are tractable and already display interesting nonequilibrium steady states~\cite{vieira2020}, many experimentally relevant platforms (for example; mesoscopic conductors, engineered quantum wires, cold-atom arrays) operate in regimes where interactions are weak but consequential. In recent work \cite{Prositto_2025}, the authors studied  a solvable model of a qubit interacting with a stream of thermalized ancilla qubits with a Heisenberg-type of Hamiltonian in the repeated-interaction protocol. Similar to the noninteracting case in Ref. \onlinecite{vieira2020}, they find steady states that are diagonal but not necessarily thermalized with the ancilla.

In this work, we will examine the quadratic resetting models of Ref. \cite{vieira2020} with weak interactions. There are two interesting questions to answer: (i) can the SPDM evolution preserve a tractable structure when interactions are included, and (ii) can such nonlinear dynamics be embedded in a completely positive (CP) Lindblad generator, yielding a continuous-time analogue of resetting? A central question is then how to incorporate interactions into the resetting picture while preserving the desirable properties of the quadratic case: (i) closure at the SPDM level, (ii) a clear connection to a microscopic open-system description, and (iii) complete positivity of the reduced dynamics. These issues connect naturally to the broader literature on quasi-free and Gaussian open fermionic systems~\cite{prosen2008,barthel2022,hackl2021,zhang2022}.

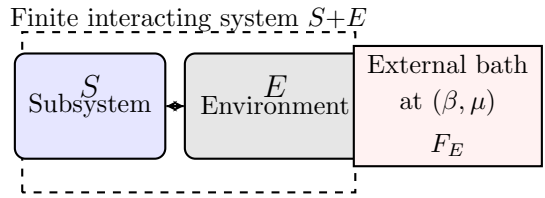
\begin{figure}[t]
\centering
\begin{tikzpicture}[scale=0.82, every node/.style={font=\large}]

\tikzstyle{blockS} = [rounded corners=4pt, thick, draw=black, fill=blue!10,
                      minimum width=2.0cm, minimum height=1.4cm]
\tikzstyle{blockE} = [rounded corners=4pt, thick, draw=black, fill=gray!20,
                      minimum width=2.4cm, minimum height=1.4cm]
\tikzstyle{bath}   = [rectangle, thick, draw=black, fill=red!5,
                      minimum width=2.5cm, minimum height=1.6cm]
\tikzstyle{arrow}  = [->, >=stealth, thick]

\draw[dashed, thick] (-3.4,-0.5) rectangle (2.0,2.1);
\node[font=\normalsize] at (-0.7,2.3) {Finite interacting system $S{+}E$};

\node[blockS] (S) at (-2.3,0.9) {};
\node at (S.north) [yshift=-0.45cm] {$S$};
\node at (S.center) [yshift=0.01cm, font=\normalsize] {Subsystem};

\node[blockE] (E) at (0.7,0.9) {};
\node at (E.north) [yshift=-0.45cm] {$E$};
\node at (E.center) [yshift=0.01cm, font=\normalsize] {Environment};

\draw[arrow] (S.east) -- (E.west);
\draw[arrow] (E.west) -- (S.east);

\node[bath] (B) at (3.5,0.9) {};
\node at (B.north) [yshift=-0.25cm, font=\normalsize] {External bath};
\node at (B.center) [yshift=0.05cm, font=\normalsize] {at $(\beta,\mu)$};
\node at (B.south) [yshift=0.30cm, font=\normalsize] {$F_E$};

\end{tikzpicture}
\caption{Three-layer architecture underlying both the resetting protocol and its CP-safe GKLS embedding.  The subsystem $S$ and environment $E$ form a finite interacting system evolving under $H_0 + H_{\mathrm{Hartree}}[V]$. Only $E$ is directly equilibrated by an external thermal bath at $(\beta,\mu)$ with target SPDM $F_E$.  The coupling to the reservoir can be realized stroboscopically as a reset $V_E \to F_E$, or continuously as Gaussian Lindblad dynamics with local jump operators $L_\alpha^\pm$ and damping matrix $\Gamma$ targeting the same $F_E$.}
\label{fig:schematic}
\end{figure}

To this end, we develop a consistent extension of resetting dynamics to weakly interacting fermionic systems.  A Hartree mean-field treatment of density-density interactions yields a nonlinear, yet closed, dynamical equation for the SPDM in which the effective one-body Hamiltonian depends self-consistently on the local occupations~\cite{Bardos2003_MF_HF,FrohlichKnowles2011_TDHF,Golse2003_MF_limit,Bonitz2016_QKT,Simenel2012_TDHF_review,MerkliBerman2012_MF_open}. To place this dynamics on a rigorous open-system footing, we construct a CP-safe Gaussian Lindblad embedding that reduces exactly to the quadratic resetting model in the noninteracting limit, and which remains completely positive when the Hartree feedback is included. Importantly, the three-layer structure of the original resetting scheme is preserved: $S$ evolves coherently while $E$ mediates the coupling to an ideal thermal reservoir, now represented either stroboscopically (via resetting) or continuously (via GKLS damping)(see Fig.~\ref{fig:schematic}). This construction builds directly on the theory of Gaussian and quasi-free Lindblad evolutions~\cite{campbell2015,diosi2014}. The present work unifies and extends these directions by showing how resetting dynamics and Gaussian Lindblad evolution can be reconciled in the presence of weak interactions. In particular, the interaction-enabled nonlinear steady states revealed here have no analogue in purely quadratic models, highlighting the importance of controlled mean-field extensions in nonequilibrium quantum settings.

We illustrate the framework in two distinct settings: (i) a ring segmentation geometry where $S$ forms the first block of a larger interacting chain, and (ii) a minimal two-site system exhibiting interaction-enabled fixed points. The numerical results show that weak interactions generate nonlinear steady states and transient structures that cannot be reproduced by tuning parameters of noninteracting GKLS or resetting models.  Our results, therefore, establish a physically consistent and mathematically controlled extension of the resetting paradigm to interacting open quantum systems.


In fermionic lattice systems described by quadratic Hamiltonians, resetting protocols have a remarkable feature: the dynamics close at the level of the single-particle density matrix (SPDM). In a typical setup, a coherent quadratic Hamiltonian acts on a ring of $N$ sites. A subset of $N_E$ sites forms an ``environment'' $E$ that is periodically reset to a fixed thermal SPDM, while the remaining $N_S = N- N_E$ sites constitute the subsystem $S$ of interest. Between resets the evolution is unitary, and after each unitary stroke the SPDM of $E$ is replaced by the target state, e.g., the diagonal Fermi-Dirac state in the repeated interaction protocol of \cite{vieira2020}. The induced dynamics of the subsystem SPDM $V_S$ is an affine map
\(V_S \mapsto A V_S A^\dagger + B,\)
where $A$ and $B$ are determined by the Hamiltonian, the partition, and the target SPDM.
This map is quasi-free and preserves the fermionic structure, and in the continuous-time limit it leads to an affine SPDM ordinary differential equation (ODE) of Lyapunov type~\cite{prosen2008,barthel2022}.

Here we address this problem by combining three ingredients. First, we review and formalize the quadratic resetting dynamics on a fermionic ring and derive the subsystem SPDM map. Second, we introduce weak density-density interactions and treat them at Hartree mean-field level. This yields a nonlinear, but closed, SPDM equation where interactions appear as a state-dependent shift of the single-particle Hamiltonian. Third, we build a CP-safe Gaussian Lindblad embedding for the resulting SPDM ODE by constructing an explicit quasi-free master equation whose covariance dynamics coincides with the desired SPDM evolution. The embedding is constructive and simple: it uses local Lindblad jump operators linear in the fermionic modes.

Our main technical result is a CP-safe Gaussian embedding theorem for SPDM equations of the form
\begin{equation}
\dot V = \frac{i}{\hbar}[V,M] - \frac12\{\Gamma,V\} + \Gamma f,
\label{eq:intro-affine}
\end{equation}
where $M$ is a Hermitian single-particle Hamiltonian, $\Gamma$ is a positive semidefinite damping matrix, and $f$ is a diagonal matrix encoding target occupations.
We show that Eq.~\eqref{eq:intro-affine} is precisely the SPDM equation induced by a Lindblad master equation with Hamiltonian
\(
\hat H_0 = \sum_{\mu\nu} a_\mu^\dagger M_{\mu\nu} a_\nu
\)
and local jump operators
\(
L_\alpha^-=\sqrt{\gamma_\alpha(1-f_\alpha)}\,a_\alpha
\)
and
\(
L_\alpha^+=\sqrt{\gamma_\alpha f_\alpha}\,a_\alpha^\dagger,
\)
where $\Gamma_{\alpha\beta}=\gamma_\alpha\delta_{\alpha\beta}$ and $f_{\alpha\beta}=f_\alpha\delta_{\alpha\beta}$.
In particular, Eq.~\eqref{eq:intro-affine} is CP by construction~\cite{lindblad1976,gorini1976,diosi2014}. We then show that replacing $M$ by a Hartree mean-field $M_{\mathrm{eff}}(V)=M+H_H(V)$, with $H_H(V)$ a diagonal functional of the SPDM, preserves complete positivity if the corresponding mean-field Hamiltonian is used in the Lindblad generator as a time-dependent Hermitian Hamiltonian.
The resulting nonlinear Hartree SPDM equation
\begin{equation}
\dot V = \frac{i}{\hbar}[V,M_{\mathrm{eff}}(V)]
- \frac12\{\Gamma,V\} + \Gamma f
\label{eq:intro-Hartree}
\end{equation}
is therefore a CP-safe mean-field Lindblad dynamics.

On the physics side, we demonstrate that interactions qualitatively change the nonequilibrium steady states of resetting dynamics.
In a ring segmentation geometry, the steady-state occupation of the subsystem acquires an interaction-dependent plateau.
In a minimal two-site model with local baths and Hartree coupling, the steady-state occupations $(n_1^{\mathrm{ss}},n_2^{\mathrm{ss}})$ as a function of $U$ trace out a nontrivial curve that cannot be obtained in any noninteracting quasi-free model with simply retuned bath parameters. This interaction-enabled Hartree fixed-point structure is absent in the noninteracting resetting model.

The rest of the paper is organized as follows. In Sec.~\ref{sec:model} we define the quadratic ring and derive the subsystem SPDM map for the resetting protocol. In Sec.~\ref{sec:Hartree-main} we introduce density-density interactions and derive the Hartree-closed SPDM evolution. Sec.~\ref{sec:CPembedding-main} contains the CP-safe Gaussian embedding theorem and its Hartree extension.
Sec.~\ref{sec:numerics-main} presents numerical results for the ring geometry and the two-site model. We conclude in Sec.~\ref{sec:discussion}. Detailed derivations of all intermediate steps are collected in Appendices~\ref{app:Hamiltonian-evolution}--\ref{app:ring-numerics}.

\section{Quadratic ring and resetting dynamics}
\label{sec:model}

In this section we set up the quadratic fermionic ring and the stroboscopic resetting protocol and derive the subsystem SPDM map that results from repeated continuation. 

\subsection{Quadratic fermionic ring}

We consider spinless fermions on a ring with $N$ sites labeled by Greek indices $\alpha,\beta,\mu,\nu\in\{1,\dots,N\}$.
The fermionic annihilation and creation operators $a_\alpha$, $a_\alpha^\dagger$ obey
\begin{equation}
\{a_\alpha,a_\beta^\dagger\}=\delta_{\alpha\beta},\qquad
\{a_\alpha,a_\beta\}=0=\{a_\alpha^\dagger,a_\beta^\dagger\}.
\label{eq:CAR}
\end{equation}
The quadratic Hamiltonian is
\begin{equation}
\hat H_0 = \sum_{\mu,\nu=1}^N a_\mu^\dagger M_{\mu\nu} a_\nu,
\qquad M=M^\dagger,
\label{eq:H0}
\end{equation}
where $M$ is the single-particle Hamiltonian matrix.
For concreteness, in the numerical examples we adopt a nearest-neighbor tight-binding Hamiltonian with on-site energies,
\begin{equation}
M_{\mu\nu} = \varepsilon_\mu \delta_{\mu\nu}
- J (\delta_{\mu,\nu+1} + \delta_{\mu,\nu-1}),
\end{equation}
with periodic boundary conditions.
However, the derivations in this and the following sections apply to an arbitrary Hermitian $M$.

The many-body density operator $\rho(t)$ evolves under the von Neumann equation,
\begin{equation}
\dot\rho(t) = -\frac{i}{\hbar}[\hat H_0,\rho(t)].
\end{equation}
We are interested in the single-particle density matrix (SPDM),
\begin{equation}
V_{\alpha\beta}(t) := \Tr\!\left[\rho(t) a_\alpha^\dagger a_\beta\right],
\qquad
V(t)\in\mathbb C^{N\times N},
\label{eq:SPDM-def}
\end{equation}
which encodes all two-point correlations for a quasi-free state.

The Heisenberg equation for the operator $O_{\alpha\beta}=a_\alpha^\dagger a_\beta$ is
\begin{equation}
\frac{d}{dt}O_{\alpha\beta}=\frac{i}{\hbar}[\hat H_0,O_{\alpha\beta}].
\end{equation}
Using the algebra of fermionic bilinears (derived explicitly in Appendix~\ref{app:Hamiltonian-evolution}) one finds
\begin{equation}
\frac{d}{dt}V_{\alpha\beta}(t)
= \frac{i}{\hbar}\sum_\mu M_{\mu\alpha} V_{\mu\beta}
 -\frac{i}{\hbar}\sum_\nu M_{\beta\nu} V_{\alpha\nu},
\end{equation}
which in matrix form is
\begin{equation}
\dot V(t)=\frac{i}{\hbar}[V(t),M].
\label{eq:SPDM-unitary}
\end{equation}
Equation~\eqref{eq:SPDM-unitary} is the basic unitary evolution equation for the SPDM under a quadratic Hamiltonian.

Formally, the solution of Eq.~\eqref{eq:SPDM-unitary} can be written in terms of the single-particle evolution operator $U_0(t)=\exp(-i M t/\hbar)$ as
\begin{equation}
V(t) = U_0(t) V(0) U_0^\dagger(t).
\label{eq:SPDM-unitary-solution}
\end{equation}
In the context of resetting, we will use Eq.~\eqref{eq:SPDM-unitary-solution} on a time interval of length $\tau$ between resets.

\subsection{Subsystem-environment partition and resetting}

We next partition the ring into a subsystem $S$ and an environment $E$ as schematically shown in Fig. 2. Without loss of generality we take $S$ to be the first $N_S$ sites and $E$ the remaining $N_E=N-N_S$ sites.
In block form, the SPDM reads
\begin{equation}
V=
\begin{pmatrix}
V_S & C\\
C^\dagger & V_E
\end{pmatrix},
\label{eq:SPDM-block}
\end{equation}
where $V_S\in\mathbb C^{N_S\times N_S}$ is the subsystem SPDM, $V_E\in\mathbb C^{N_E\times N_E}$ the environment SPDM, and $C$ the system--environment correlation block.
We also write the single-particle Hamiltonian $M$ in the same block form,
\begin{equation}
M =
\begin{pmatrix}
M_{SS} & M_{SE}\\
M_{ES} & M_{EE}
\end{pmatrix}.
\end{equation}

\begin{figure}[t]
    \centering
 {\includegraphics[width=0.48\textwidth]{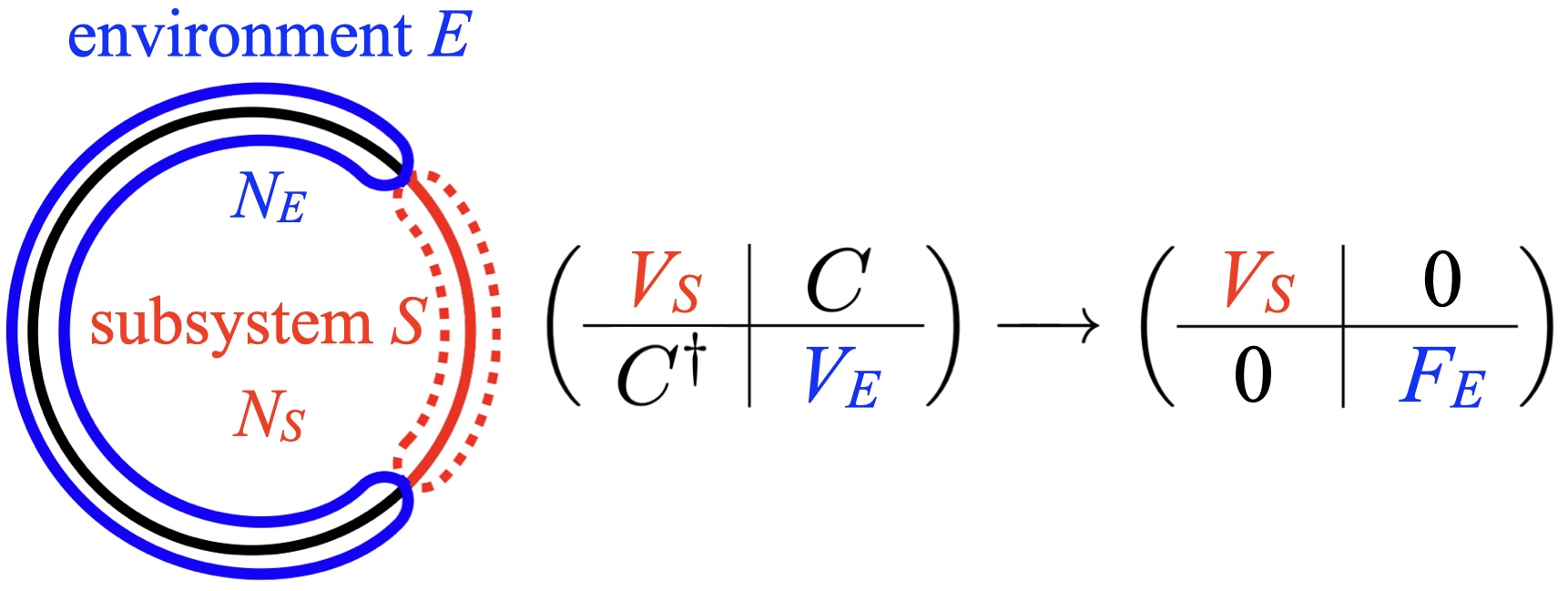}} 
  \caption{A schematic of the quadratic fermionic ring comprising $N_E$ sites in the environment (blue) and $N_S$ in the subsystem (red), with $N = N_S + N_E$ total sites. At each reset step the subsystem SPDM block $V_S$ is left intact, the environment block $V_E$ is set to the target $F_E$, and the coherences are reset to zero. This corresponds to the RI protocol of Ref. \onlinecite{vieira2020}. See text for details.}
\end{figure}

The resetting protocol proceeds in discrete steps of duration $\tau$:

\begin{enumerate}
\item Immediately before a reset the SPDM is $V^{(-)}$, written as in Eq.~\eqref{eq:SPDM-block}.
\item At the reset, the environment block is replaced by a fixed target SPDM $F_E$ and all correlations between $S$ and $E$ are erased:
\begin{equation}
V^{(0)}=
\begin{pmatrix}
V_S & 0\\
0   & F_E
\end{pmatrix}.
\label{eq:reset-state}
\end{equation}
\item Between resets, the ring evolves unitarily for time $\tau$ under $\hat H_0$. Using Eq.~\eqref{eq:SPDM-unitary-solution}, the SPDM immediately after the unitary evolution is
\begin{equation}
V^{(+)} = U_0 V^{(0)} U_0^\dagger,
\qquad
U_0 = e^{-\tfrac{i}{\hbar}M\tau}.
\label{eq:SPDM-after-unitary}
\end{equation}
\item The updated subsystem SPDM is extracted as the $S\times S$ block of $V^{(+)}$:
\begin{equation}
V_S' = \left[V^{(+)}\right]_{SS}.
\end{equation}
\end{enumerate}

Writing the single-particle unitary $U_0$ in block form,
\begin{equation}
U_0 =
\begin{pmatrix}
U_{SS} & U_{SE}\\
U_{ES} & U_{EE}
\end{pmatrix},
\end{equation}
we insert Eq.~\eqref{eq:reset-state} into Eq.~\eqref{eq:SPDM-after-unitary} and obtain
\begin{align}
V^{(+)}
&= 
\begin{pmatrix}
U_{SS} & U_{SE}\\
U_{ES} & U_{EE}
\end{pmatrix}
\begin{pmatrix}
V_S & 0\\
0   & F_E
\end{pmatrix}
\begin{pmatrix}
U_{SS}^\dagger & U_{ES}^\dagger\\
U_{SE}^\dagger & U_{EE}^\dagger
\end{pmatrix}
\nonumber\\[2mm]
&=
\begin{pmatrix}
U_{SS} V_S U_{SS}^\dagger + U_{SE} F_E U_{ES}^\dagger
&
\cdots
\\
\cdots & \cdots
\end{pmatrix}.
\end{align}
The updated subsystem SPDM is therefore
\begin{equation}
V_S' = U_{SS} V_S U_{SS}^\dagger + U_{SE} F_E U_{ES}^\dagger.
\label{eq:reset-map-main}
\end{equation}
This is the central subsystem map generated by one cycle of unitary evolution followed by a reset.

Equation~\eqref{eq:reset-map-main} has two distinct pieces: a homogeneous part $V_S\mapsto U_{SS}V_S U_{SS}^\dagger$ and a constant source term
\(
B:=U_{SE}F_E U_{ES}^\dagger
\).
Both are completely determined by the full unitary $U_0$ and the environment target state $F_E$. Importantly, Eq.~\eqref{eq:reset-map-main} is linear and quasi-free at the SPDM level.

In the following sections we regard Eq.~\eqref{eq:reset-map-main} as the discrete-time reference dynamics and build a continuous-time description that retains its affine structure while providing a transparent connection to GKLS master equations.

\section{Weak interactions and Hartree closure}
\label{sec:Hartree-main}
To make contact with interacting fermionic systems we now add density-density interactions to the quadratic Hamiltonian and derive a mean-field SPDM equation that closes at the single-particle level.

\subsection{density-density interactions}

We consider fermions on $N$ sites with a quadratic Hamiltonian and a two-body density-density interaction,
\begin{equation}
  \hat H = \hat H_0 + \hat H_{\mathrm{int}},
  \qquad
  \hat H_0 = \sum_{\mu,\nu=1}^N M_{\mu\nu}\,\hat a_\mu^\dagger \hat a_\nu,
\end{equation}
with $M_{\mu\nu} = M_{\nu\mu}^*$, and
\begin{equation}
  \hat H_{\mathrm{int}}
  = \frac12 \sum_{\gamma,\delta=1}^N
      U_{\gamma\delta}\,\hat n_\gamma \hat n_\delta,
  \qquad
  \hat n_\gamma = \hat a_\gamma^\dagger \hat a_\gamma,
  \label{eq:Hint-main}
\end{equation}
where $U_{\gamma\delta} = U_{\delta\gamma}\in\mathbb{R}$ is a symmetric interaction matrix (e.g., nearest-neighbour density-density
interactions on a chain or ring~\cite{KunduSenechal2024}).  We continue to define the single-particle density matrix (SPDM) as
\begin{equation}
  V_{\alpha\beta}(t)
  = \big\langle \hat a_\alpha^\dagger \hat a_\beta \big\rangle_t
  = \mathrm{Tr}\big[\hat\rho(t)\,\hat a_\alpha^\dagger \hat a_\beta\big].
\end{equation}

In the absence of interactions, the Hamiltonian contribution to the SPDM dynamics has the well-known form
\begin{equation}
  \dot V\Big|_{H_0} = \frac{i}{\hbar}[V,M],
  \label{eq:Vdot-quadratic}
\end{equation}
where $M$ is the single-particle Hamiltonian entering the quadratic von Neumann equation, and hence
$V(t) = e^{-iMt/\hbar} V(0)\,e^{+iMt/\hbar}$.

To incorporate interactions while retaining a closed SPDM dynamics, we use a Hartree mean-field approximation.  At the operator level we perform the standard decoupling
\begin{equation}
  \hat n_\gamma \hat n_\delta
  \approx \hat n_\gamma\,\langle \hat n_\delta\rangle
        + \langle \hat n_\gamma\rangle\,\hat n_\delta
        - \langle \hat n_\gamma\rangle\langle \hat n_\delta\rangle,
\end{equation}
and drop the $c$-number term
$\langle \hat n_\gamma\rangle\langle \hat n_\delta\rangle$, which
only shifts the total energy.  Writing
\begin{equation}
  n_\gamma(t) \equiv \langle \hat n_\gamma\rangle_t
  = \langle \hat a_\gamma^\dagger \hat a_\gamma\rangle_t
  = V_{\gamma\gamma}(t),
\end{equation}
we obtain the Hartree mean-field Hamiltonian
\begin{equation}
  \hat H_{\mathrm{H}}[V]
  = \sum_{\gamma}
      h_\gamma(V)\,\hat n_\gamma,
  \quad
  h_\gamma(V) = \sum_{\delta} U_{\gamma\delta}\,V_{\delta\delta}.
  \label{eq:Hartree-potential}
\end{equation}
Thus $\hat H_{\mathrm{int}}$ is replaced, at Hartree level, by the effective quadratic Hamiltonian $\hat H_{\mathrm{H}}[V]$ plus an
irrelevant constant, and the total Hamiltonian in the von Neumann equation is approximated by
\begin{equation}
  \hat H_{\mathrm{eff}}[V]
  = \hat H_0 + \hat H_{\mathrm{H}}[V].
\end{equation}

Using the general relation
\begin{equation}
  i\hbar \,\frac{d}{dt}\langle \hat O\rangle_t
  = \big\langle [\hat O,\hat H_{\mathrm{eff}}[V]]\big\rangle_t,
\end{equation}
we now compute the Hartree contribution to the SPDM.  For $\hat O_{\alpha\beta} = \hat a_\alpha^\dagger \hat a_\beta$ we have
\begin{align}
  i\hbar\,\dot V_{\alpha\beta}\Big|_{\mathrm{H}}
  &= \big\langle
        [\hat a_\alpha^\dagger \hat a_\beta,\hat H_{\mathrm{H}}[V]]
      \big\rangle \notag\\
   &= \sum_{\gamma} h_\gamma(V)\,
      \big\langle
        [\hat a_\alpha^\dagger \hat a_\beta,\hat n_\gamma]
      \big\rangle.
\end{align}
Using $\hat n_\gamma = \hat a_\gamma^\dagger \hat a_\gamma$ and the canonical anti\-commutation relations, one finds
\begin{equation}
  [\hat n_\gamma,\hat a_\alpha^\dagger]
  = \delta_{\gamma\alpha}\,\hat a_\alpha^\dagger,
  \qquad
  [\hat n_\gamma,\hat a_\beta]
  = -\delta_{\gamma\beta}\,\hat a_\beta,
\end{equation}
and hence
\begin{align}
  [\hat n_\gamma,\hat a_\alpha^\dagger \hat a_\beta]
  &= [\hat n_\gamma,\hat a_\alpha^\dagger]\hat a_\beta
   + \hat a_\alpha^\dagger[\hat n_\gamma,\hat a_\beta]\notag\\
   &= (\delta_{\gamma\alpha} - \delta_{\gamma\beta})
     \hat a_\alpha^\dagger \hat a_\beta,\\
  [\hat a_\alpha^\dagger \hat a_\beta,\hat n_\gamma]
  &= -[\hat n_\gamma,\hat a_\alpha^\dagger \hat a_\beta]
   = (\delta_{\gamma\beta} - \delta_{\gamma\alpha})
      \hat a_\alpha^\dagger \hat a_\beta.
\end{align}
Therefore
\begin{align}
  [\hat a_\alpha^\dagger \hat a_\beta,\hat H_{\mathrm{H}}[V]]
  &= \sum_\gamma h_\gamma(V)\,
     (\delta_{\gamma\beta} - \delta_{\gamma\alpha})
      \hat a_\alpha^\dagger \hat a_\beta\\
  &= \big(h_\beta(V) - h_\alpha(V)\big)\,
      \hat a_\alpha^\dagger \hat a_\beta,
\end{align}
and taking expectation values yields
\begin{equation}
  i\hbar\,\dot V_{\alpha\beta}\Big|_{\mathrm{H}}
  = \big(h_\beta(V) - h_\alpha(V)\big)\,V_{\alpha\beta}.
  \label{eq:Hartree-component}
\end{equation}

It is convenient to package the Hartree term into a diagonal single-particle matrix $M_{\mathrm{H}}(V)$ acting on the SPDM.  With
our convention $V_{\alpha\beta} = \langle\hat a_\alpha^\dagger \hat a_\beta\rangle$ and the quadratic evolution
$\dot V|_{H_0} = (i/\hbar)[V,M]$ in Eq.~\eqref{eq:Vdot-quadratic}, a
direct calculation shows that the commutator
\begin{equation}
  \big[V,M_{\mathrm{H}}(V)\big]_{\alpha\beta}
  = \big(h_\beta(V) - h_\alpha(V)\big)\,V_{\alpha\beta}
\end{equation}
is obtained by defining
\begin{equation}
  \big(M_{\mathrm{H}}(V)\big)_{\alpha\beta}
  = h_\beta(V)\,\delta_{\alpha\beta}.
  \label{eq:Hartree-M-matrix}
\end{equation}
Comparing with Eq.~\eqref{eq:Hartree-component}, we can therefore write the Hamiltonian part of the SPDM dynamics in the compact form
\begin{equation}
  \dot V\Big|_{\mathrm{Ham}}
  = \frac{i}{\hbar}[V,M] + \frac{i}{\hbar}[V,M_{\mathrm{H}}(V)]
  = \frac{i}{\hbar}\big[V,M_{\mathrm{eff}}(V)\big],
  \label{eq:SPDM-Hartree-closed}
\end{equation}
with the effective single-particle Hamiltonian
\begin{equation}
  M_{\mathrm{eff}}(V) = M + M_{\mathrm{H}}(V).
  \label{eq:Meff-def}
\end{equation}
Equation~\eqref{eq:SPDM-Hartree-closed} is a nonlinear, yet closed, equation for the SPDM: the Hartree potential depends only on the
diagonal occupations $n_\gamma(t) = V_{\gamma\gamma}(t)$.  For a nearest-neighbour density-density interaction on a ring,
$U_{\gamma\delta} = U\,\mathrm{Adj}_{\gamma\delta}$, we have
\begin{equation}
  h_\gamma(V) = U \sum_{\delta}\mathrm{Adj}_{\gamma\delta}
  V_{\delta\delta},\quad
  \big(M_{\mathrm{H}}(V)\big)_{\alpha\beta}
  = h_\beta(V)\,\delta_{\alpha\beta},
\end{equation}
and hence $M_{\mathrm{eff}}(V)$ differs from the quadratic single-particle Hamiltonian $M$ by a diagonal, occupation-dependent
Hartree shift on each site.

\subsection{Hartree-extended resetting and continuous-time limit}

In the resetting protocol, the unitary evolution between resets is now generated by the mean-field Hamiltonian $\hat H_{\mathrm{eff}}[V]$ with single-particle matrix $M_{\mathrm{eff}}(V)$. For sufficiently weak interactions and short reset periods $\tau$ one can treat $M_{\mathrm{eff}}(V)$ as approximately constant during a single interval and use Eq.~\eqref{eq:SPDM-unitary-solution} with $M\to M_{\mathrm{eff}}(V)$. As in the noninteracting case, this leads to a subsystem map of the form of Eq.~\eqref{eq:reset-map-main}, but with $U_0$ computed from $M_{\mathrm{eff}}(V)$.

In the limit of frequent resets, the discrete map converges to a continuous-time description where the homogeneous part becomes a commutator with an effective Hamiltonian and the inhomogeneous part becomes a linear dissipator with a constant source term.
Motivated by this structure, in the next section we construct a continuous-time SPDM equation with the general affine form of Eq.~\eqref{eq:intro-Hartree} and show how it can be embedded into a CP Gaussian Lindblad dynamics.

\section{CP-safe Gaussian embedding}
\label{sec:CPembedding-main}

We now construct an explicit fermionic Lindblad master equation whose SPDM dynamics coincides with an affine ODE of the form~\eqref{eq:intro-affine} and prove that this constitutes a completely positive (CP) Gaussian embedding.
We then show how to extend this construction to the Hartree case.

\subsection{Lindblad model and SPDM equation}

We consider the Lindblad master equation
\begin{equation}
\dot\rho = -\frac{i}{\hbar}[\hat H_0,\rho]
+ \sum_{\alpha=1}^N
\left(\mathcal D[L_\alpha^-](\rho) + \mathcal D[L_\alpha^+](\rho)\right),
\label{eq:master-CP}
\end{equation}
with Hamiltonian $\hat H_0$ given by Eq.~\eqref{eq:H0} and local jump operators
\begin{equation}
L_\alpha^- = \sqrt{\gamma_\alpha(1-f_\alpha)}\,a_\alpha,\qquad
L_\alpha^+ = \sqrt{\gamma_\alpha f_\alpha}\,a_\alpha^\dagger.
\label{eq:Lops-main}
\end{equation}
Here $\gamma_\alpha\ge 0$ are damping rates and $f_\alpha\in[0,1]$ encode target occupations.
The dissipator is
\begin{equation}
\mathcal D[L](\rho) = L\rho L^\dagger
-\frac12\left(L^\dagger L\rho + \rho L^\dagger L\right).
\end{equation}
Equation~\eqref{eq:master-CP} is manifestly in GKLS form and hence generates a completely positive, trace-preserving (CPTP) evolution of $\rho(t)$ for all $t\ge 0$.

We now derive the induced equation of motion for the SPDM
$V_{\mu\nu}(t)=\Tr[\rho(t)a_\mu^\dagger a_\nu]$.
The Hamiltonian contribution is the same as in Sec.~\ref{sec:model} and yields $\dot V|_H=\tfrac{i}{\hbar}[V,M]$.
The dissipative part can be computed by evaluating
\(
\Tr(\mathcal D[L_\alpha^\pm](\rho) a_\mu^\dagger a_\nu)
\)
and using cyclicity of the trace and the canonical anticommutation relations.
A detailed step-by-step derivation is provided in Appendix~\ref{app:dissipator-details}.
The result can be summarized as follows.

Define diagonal matrices
\begin{equation}
\Gamma_{\alpha\beta} := \gamma_\alpha\delta_{\alpha\beta},\qquad
f_{\alpha\beta} := f_\alpha\delta_{\alpha\beta}.
\end{equation}
Then the dissipative contribution to the SPDM is
\begin{equation}
\dot V\Big|_{\mathrm{diss}}
= -\frac12\{\Gamma,V\} + \Gamma f.
\label{eq:Vdot-diss-main}
\end{equation}
For diagonal entries $V_{\alpha\alpha}$ this reduces to the single-mode equation
\begin{equation}
\dot V_{\alpha\alpha} = -\gamma_\alpha(V_{\alpha\alpha}-f_\alpha),
\end{equation}
i.e., an exponential relaxation towards the target occupation $f_\alpha$.
For off-diagonal entries $V_{\alpha\beta}$ with $\alpha\neq\beta$ one obtains
\begin{equation}
\dot V_{\alpha\beta} = -\frac12(\gamma_\alpha+\gamma_\beta)\,V_{\alpha\beta},
\end{equation}
i.e., exponential decoherence without a source term.

Combining Hamiltonian and dissipative parts gives the affine SPDM ODE
\begin{equation}
\dot V = \frac{i}{\hbar}[V,M] - \frac12\{\Gamma,V\} + \Gamma f,
\label{eq:SPDM-affine-main}
\end{equation}
which matches the general structure anticipated in Eq.~\eqref{eq:intro-affine}. The affine SPDM equation extends to the Hartree-closed equation
\begin{equation}
  \dot V
  = \frac{i}{\hbar}[V,M_{\mathrm{eff}}(V)]
    - \frac12\{\Gamma,V\}
    + \Gamma f,
  \label{eq:SPDM-Hartree-CP-main}
\end{equation}
with $M_{\mathrm{eff}}(V) = M + M_{\mathrm{H}}(V)$.  Throughout this section we assume that
$M=M^\dagger$ is Hermitian, that the damping matrix $\Gamma_{\alpha\beta} = \gamma_\alpha \delta_{\alpha\beta}$ is
positive semidefinite with $\gamma_\alpha\ge 0$, and that $f_{\alpha\beta} = f_\alpha \delta_{\alpha\beta}$ with
$0\le f_\alpha\le 1$. 

\subsection{Embedding theorem}

\noindent\textbf{Theorem 1 (CP-safe Gaussian embedding).}
\emph{
Let $M=M^\dagger\in\mathbb C^{N\times N}$, $\Gamma=\mathrm{diag}(\gamma_1,\dots,\gamma_N)$ with $\gamma_\alpha\ge 0$, and $f=\mathrm{diag}(f_1,\dots,f_N)$ with $0\le f_\alpha\le 1$.
Consider the fermionic Lindblad master equation~\eqref{eq:master-CP} with Hamiltonian $\hat H_0$ given by Eq.~\eqref{eq:H0} and Lindblad operators $L_\alpha^\pm$ given by Eq.~\eqref{eq:Lops-main}.
Let $V(t)$ denote the SPDM with entries $V_{\mu\nu}(t)=\Tr[\rho(t)a_\mu^\dagger a_\nu]$.
Then $V(t)$ obeys the affine SPDM equation~\eqref{eq:SPDM-affine-main} for all $t\ge 0$.
In particular, the SPDM dynamics generated by Eq.~\eqref{eq:SPDM-affine-main} is CP and trace preserving.
}
\medskip

\noindent\emph{Proof.}
The proof is constructive and proceeds by direct evaluation.
We first compute the Hamiltonian contribution, which yields Eq.~\eqref{eq:SPDM-unitary} and hence $\dot V|_H=\tfrac{i}{\hbar}[V,M]$ (Appendix~\ref{app:Hamiltonian-evolution}).
We then evaluate the dissipative contribution using the explicit form of $L_\alpha^\pm$ and the canonical anticommutation relations.
For a single mode the calculation reduces to that of Appendix~\ref{app:dissipator-details} and gives $\dot n = -\gamma(n-f)$.
For many modes, each diagonal SPDM element $V_{\alpha\alpha}$ obeys the same single-mode equation with its own rate $\gamma_\alpha$ and target $f_\alpha$, while off-diagonal elements decay at rate $(\gamma_\alpha+\gamma_\beta)/2$ without a source.
Collecting these results in matrix form yields Eq.~\eqref{eq:Vdot-diss-main}.
Adding the Hamiltonian and dissipative parts yields Eq.~\eqref{eq:SPDM-affine-main}, which is therefore equivalent to the SPDM dynamics induced by the GKLS master equation~\eqref{eq:master-CP}.
Since Eq.~\eqref{eq:master-CP} is completely positive and trace preserving by construction, so is Eq.~\eqref{eq:SPDM-affine-main}.

\subsection{Hartree extension and CP safety}

We now extend the Gaussian embedding to include the Hartree mean-field $H_H(V)$.
Define the effective single-particle Hamiltonian
\begin{equation}
M_{\mathrm{eff}}(V) := M + H_H(V),
\end{equation}
and the corresponding quadratic Hamiltonian
\begin{equation}
\hat H_{\mathrm{eff}}[V(t)] = \sum_{\mu\nu} a_\mu^\dagger
M_{\mathrm{eff}}(V(t))_{\mu\nu} a_\nu.
\end{equation}
We consider the time-local master equation
\begin{align}
\dot\rho(t)
&= -\frac{i}{\hbar}[\hat H_{\mathrm{eff}}[V(t)],\rho(t)] \nonumber\\
&+ \sum_\alpha\left(\mathcal D[L_\alpha^-](\rho(t))
+ \mathcal D[L_\alpha^+](\rho(t))\right),
\label{eq:master-Hartree}
\end{align}
with the same local jump operators $L_\alpha^\pm$ as before.

For each fixed time $t$ this generator has the standard GKLS structure with a Hermitian Hamiltonian and positive rates.
The evolution from time $t$ to time $t+\mathrm dt$ is therefore a CP, trace-preserving map.
The full evolution from time $0$ to time $T$ is a time-ordered composition of such maps and is thus CP and trace preserving.

Taking SPDM matrix elements and repeating the derivation of Sec.~\ref{sec:CPembedding-main}, but with $M$ replaced by $M_{\mathrm{eff}}(V)$ in the Hamiltonian part, yields the nonlinear SPDM equation
\begin{equation}
\dot V = \frac{i}{\hbar}[V,M_{\mathrm{eff}}(V)]
- \frac12\{\Gamma,V\} + \Gamma f.
\label{eq:SPDM-Hartree-CP-main}
\end{equation}
The dissipative part is unchanged and retains the same CP properties as in Theorem~1.
The only nonlinearity comes from the Hartree functional $H_H(V)$ in the commutator.
Since $\hat H_{\mathrm{eff}}[V(t)]$ is Hermitian for all $V(t)$, the evolution generated by Eq.~\eqref{eq:master-Hartree} remains CP-safe.

Equation~\eqref{eq:SPDM-Hartree-CP-main} is the central mean-field SPDM equation we use in the following section.
It has the same affine structure as the quadratic resetting dynamics, but with a state-dependent effective Hamiltonian.

\section{Interaction-enabled steady states: numerical examples}
\label{sec:numerics-main}

We now illustrate the CP-safe Hartree dynamics in two complementary geometries.
In both cases we emphasize how interactions modify the nonequilibrium steady state in a way that is absent in the noninteracting resetting model.

\subsection{Ring segmentation: resetting versus GKLS embedding}
\begin{figure}[t]
  \centering
  \includegraphics[width=\linewidth]{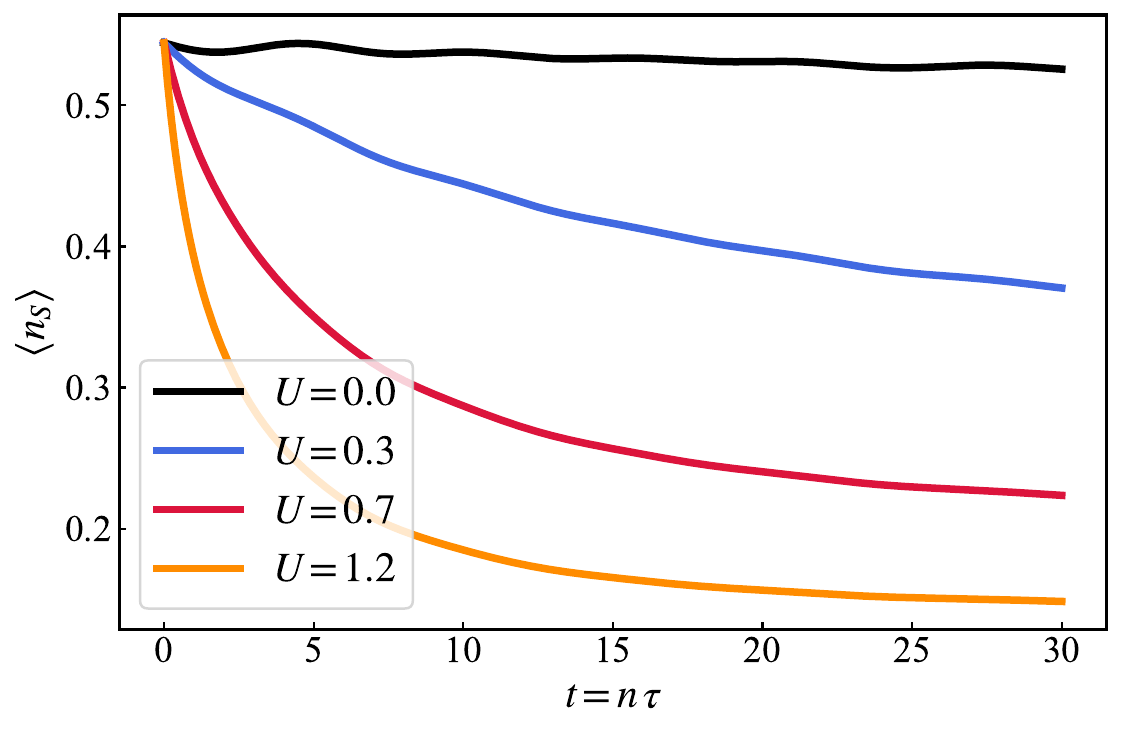}
  \caption{Time evolution of the average subsystem occupation $\langle n_S(t)\rangle$ for several interaction strengths $U$ in the resetting protocol. The ring has $N = N_S + N_E = 110$ sites with $N_S = 10$ subsystem sites and $N_E = 100$ environment sites.
  At each stroboscopic step of duration $\tau$ the environment SPDM is reset to a thermal target $F_E$ while the subsystem block is left unchanged. The mean-field Hamiltonian includes an on-site Hartree shift proportional to the local occupation. As $U$ increases, the steady-state plateau of $\langle n_S(t)\rangle$ shifts, reflecting the Hartree feedback between the subsystem and the environment. The approach to the plateau is smooth and monotone in time.}
  \label{fig:ring-RI}
\end{figure}
\begin{figure}[t]
  \centering
  \includegraphics[width=\linewidth]{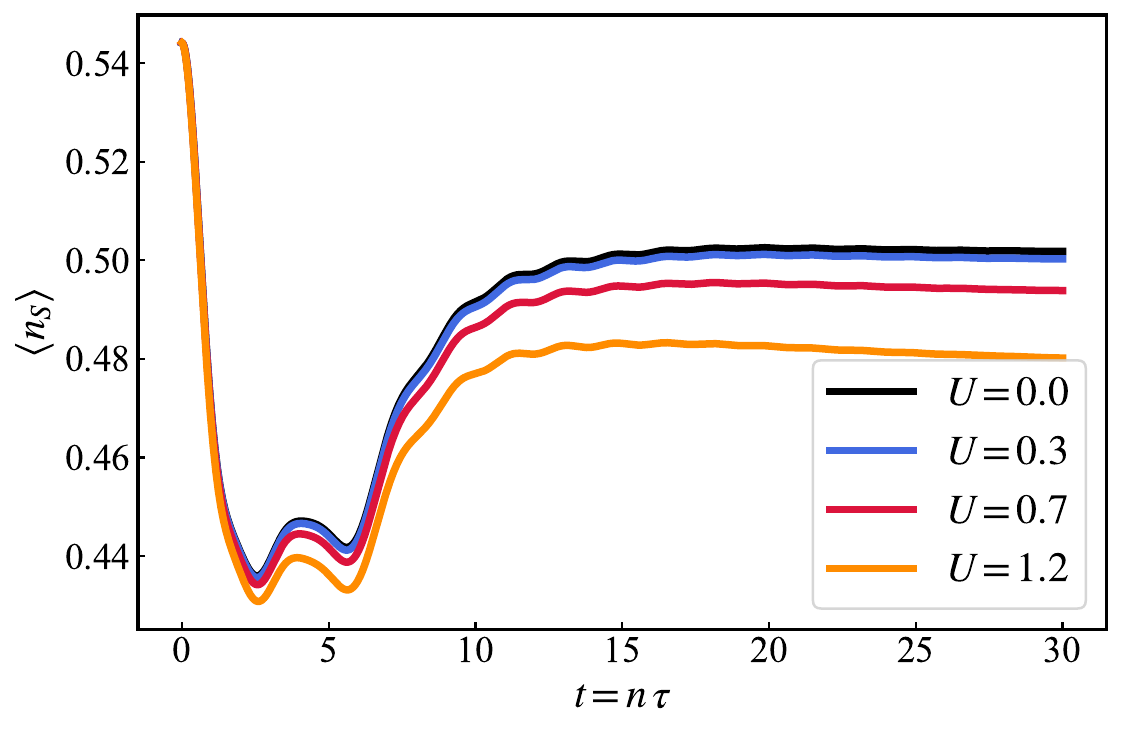}
  \caption{Time evolution of the average subsystem occupation $\langle n_S(t)\rangle$ under the CP-safe Hartree SPDM equation~\eqref{eq:SPDM-Hartree-CP-main} for the same ring geometry and parameters as in Fig.~\ref{fig:ring-RI}. The environment sites experience local Lindblad damping with rate $\gamma$ towards the same thermal target $F_E$ used in the resetting protocol, while the subsystem sites are undamped ($\gamma_\alpha = 0$ on $S$). For the parameter choice used here, the long-time values of $\langle n_S(t)\rangle$ are close to, but not exactly identical with, the plateaus of the resetting dynamics, since the discrete full reset and the continuous GKLS damping realise different bath couplings. Nevertheless, the dependence of the steady-state plateau on $U$ and the direction of the Hartree-induced shift agree with Fig.~\ref{fig:ring-RI}, showing that the GKLS dynamics provides a CP-safe continuous-time analogue of the Hartree-extended resetting model. The continuous-time evolution exhibits damped oscillations superimposed on the overall relaxation, a generic feature of coherent dynamics in the presence of Markovian dissipation.}
  \label{fig:ring-GKLS}
\end{figure}

We first return to the segmentation of a fermionic ring into a subsystem $S$ and environment $E$.
We choose a ring with $N=N_S+N_E$ sites, nearest-neighbor hopping $J$, uniform on-site energies, and on-site Hartree interactions $U_{\gamma\delta}=U\delta_{\gamma\delta}$.
We consider a subsystem $S$ of $N_S=10$ contiguous sites and an environment $E$ of $N_E=100$ sites.
The environment is repeatedly reset to a thermal (diagonal Fermi-Dirac) SPDM $F_E$ at inverse temperature $\beta$ and chemical potential $\mu$, while the subsystem is left untouched at the reset.

In the resetting protocol we iterate the Hartree-extended map of Eq.~\eqref{eq:reset-map-main}, updating the mean-field Hamiltonian between steps.
In the GKLS embedding we evolve the full SPDM under Eq.~\eqref{eq:SPDM-Hartree-CP-main} with a diagonal damping matrix $\Gamma$ that is nonzero only on environment sites and a target $f$ that coincides with $F_E$ on $E$ and vanishes on $S$.
The initial state is chosen such that the environment SPDM equals $F_E$ and the subsystem SPDM has an occupation profile out of equilibrium with respect to the environment.

We monitor the average subsystem occupation
\begin{equation}
  \langle n_S(t)\rangle = \frac{1}{N_S}\sum_{\alpha\in S} V_{\alpha\alpha}(t)
  \label{eq:nS-def}
\end{equation}
for different interaction strengths $U$. Figure~\ref{fig:ring-RI} shows $\langle n_S(t)\rangle$ under the resetting dynamics.  For each $U$ the subsystem relaxes smoothly and monotonically towards a plateau value $\bar n_S^{\mathrm{RI}}(U)$ that depends on $U$ through the Hartree feedback between $S$ and $E$.  The monotone approach reflects the stroboscopic nature of the protocol: at the beginning of each stroke the environment block is exactly reinitialized to $F_E$, so the subsystem is driven by a sequence of identical CPTP maps $\Phi_\tau^{\mathrm{RI}}$ generated by unitary evolution under $H_0+H_{\mathrm{Hartree}}[V]$ followed by a full reset on $E$. The plateau $\bar n_S^{\mathrm{RI}}(U)$ is the fixed point of this discrete map.

In contrast, Fig.~\ref{fig:ring-GKLS} shows the dynamics of $\langle n_S(t)\rangle$ under the GKLS Hartree equation~(45). The
same interaction–dependent plateaus are reached, but the transients display damped oscillations characteristic of continuous-time Lindblad evolution. The long-time values $\bar n_S^{\mathrm{GKLS}}(U)$ are close to, but not exactly identical with, the resetting plateaus $\bar n_S^{\mathrm{RI}}(U)$: the curves in Figs.~\ref{fig:ring-RI} and \ref{fig:ring-GKLS} differ at the level
of a few percent.  This small systematic offset is expected, because the two models realise different bath couplings.  The resetting
protocol implements a stroboscopic map \( V \mapsto \Phi_\tau^{\mathrm{RI}}[V],\) whereas the GKLS dynamics solves the continuous-time affine SPDM equation (Eq. (\ref{eq:SPDM-Hartree-CP-main})). In the quadratic case one can define an effective generator $\mathcal{L}_{\mathrm{RI}}$ via $\Phi_\tau^{\mathrm{RI}} = \exp(\tau \mathcal{L}_{\mathrm{RI}})$ and $\mathcal{L}_{\mathrm{GKLS}}$ via the right-hand side of the equation above.  The steady states coincide only if the rates are calibrated so that $\mathcal{L}_{\mathrm{GKLS}} = \mathcal{L}_{\mathrm{RI}}$, which would enforce $\bar n_S^{\mathrm{GKLS}}(U) = \bar n_S^{\mathrm{RI}}(U)$ for all $U$.  In the present work we do not perform this fine tuning; instead, $\tau$ and the Lindblad rate $\gamma$ are chosen independently, so a small mismatch between $\bar n_S^{\mathrm{GKLS}}(U)$ and $\bar n_S^{\mathrm{RI}}(U)$ is natural.  Importantly, both families of plateaus exhibit the same systematic Hartree-induced shift with $U$, and the GKLS dynamics shares the same affine SPDM structure as the resetting map. This shows that the CP-safe GKLS equation provides a faithful continuous-time analogue of the Hartree-extended resetting model in the large-bath limit, even though the two steady states are not exactly equal for the specific parameter choice used in the figures.

\subsection{Two-site GKLS+Hartree model: nonlinear steady states}

To highlight the role of interactions at minimal cost, we consider a two-site system with Hamiltonian
\begin{equation}
M =
\begin{pmatrix}
\varepsilon_1 & -J\\
-J & \varepsilon_2
\end{pmatrix},
\end{equation}
local baths with rates $\gamma_1=\gamma_2=\gamma$ and target occupations $f_1,f_2$, and Hartree shifts
\begin{equation}
h_1(V)=U V_{22},\qquad h_2(V)=U V_{11}.
\end{equation}
The effective single-particle Hamiltonian reads
\begin{equation}
M_{\mathrm{eff}}(V) =
\begin{pmatrix}
\varepsilon_1 + U V_{22} & -J\\
-J & \varepsilon_2 + U V_{11}
\end{pmatrix}.
\end{equation}
We parametrize the SPDM as
\begin{equation}
V =
\begin{pmatrix}
n_1 & c\\
c^* & n_2
\end{pmatrix},
\end{equation}
with occupations $n_1,n_2$ and coherence $c$.
Inserting this into Eq.~\eqref{eq:SPDM-Hartree-CP-main} and equating matrix elements yields the coupled nonlinear equations (Appendix~\ref{app:two-site-equations})
\begin{align}
\dot n_1
&= \frac{2J}{\hbar}\,\Im c - \gamma_1(n_1-f_1),\label{eq:n1-eq}\\
\dot n_2
&= -\frac{2J}{\hbar}\,\Im c - \gamma_2(n_2-f_2),\label{eq:n2-eq}\\
&\dot c
= \frac{i}{\hbar}\big[(\varepsilon_2 + U n_1)-(\varepsilon_1 + U n_2)\big]\,c \notag\\
  &+ \frac{iJ}{\hbar}(n_2-n_1)
  - \frac{1}{2}(\gamma_1+\gamma_2)c. \label{eq:c-eq}
\end{align}

We integrate Eqs.~\eqref{eq:n1-eq}--\eqref{eq:c-eq} numerically using a simple Euler method with time step $\Delta t$.
After each step we reconstruct the SPDM matrix $V$, symmetrize it to enforce Hermiticity, diagonalize it, clip its eigenvalues to the physical interval $[0,1]$, and reconstruct $V$; this guarantees that the numerical SPDM remains a valid fermionic covariance matrix at all times.

\begin{figure}[t]
  \centering
  \includegraphics[width=\linewidth]{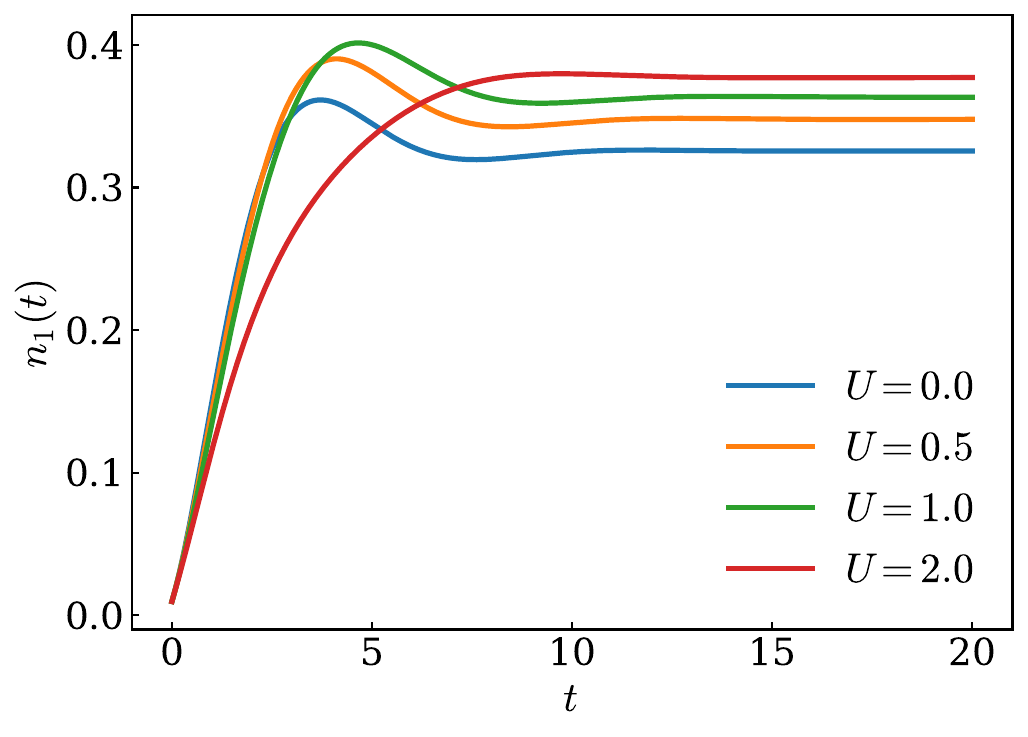}
  \caption{Time evolution of the occupation $n_1(t)=V_{11}(t)$ for the two-site GKLS+Hartree model with interaction strengths
  $U=0, 0.5, 1.0, 2.0$ (from bottom to top at long times).
  The parameters are $\varepsilon_1=0$, $\varepsilon_2=0.5$, $J=0.3$, $\gamma_1=\gamma_2=0.5$, and $f_1=0.2$, $f_2=0.8$.
  For $U=0$ the system relaxes to a noninteracting steady state governed by the quadratic GKLS Liouvillian.
  As $U$ increases, the Hartree feedback modifies both the transient dynamics and the steady-state value $n_1^{\mathrm{ss}}(U)$, as quantified in Fig. \ref{fig:two-site-nss}.}
  \label{fig:two-site-n1t}
\end{figure}

\begin{figure}[t]
  \centering
  \includegraphics[width=\linewidth]{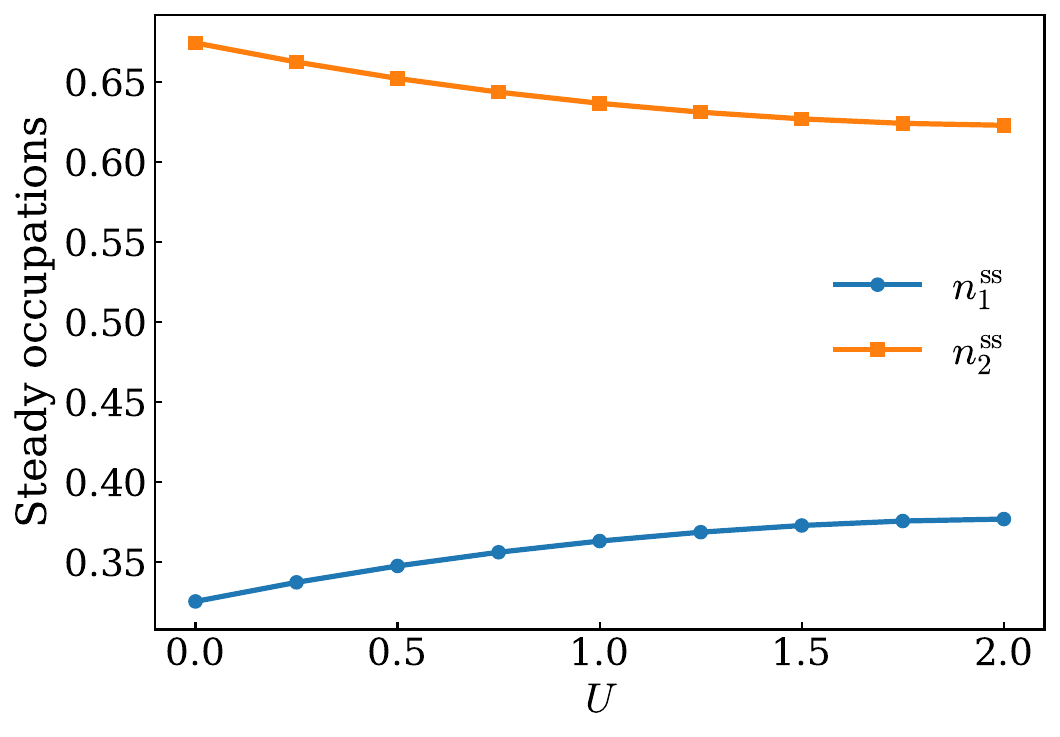}
  \caption{Steady-state occupations $n_1^{\mathrm{ss}}(U)$ and $n_2^{\mathrm{ss}}(U)$ extracted from the long-time limit of the two-site GKLS+Hartree dynamics as functions of the interaction strength $U$.
  The parameters are the same as in Fig.~\ref{fig:two-site-n1t}.
  Both occupations depend monotonically on $U$ and move in opposite directions.
  The resulting curve $(n_1^{\mathrm{ss}}(U),n_2^{\mathrm{ss}}(U))$ in the $(n_1,n_2)$ plane represents a family of nonlinear Hartree fixed points.
  Such an interaction-enabled steady-state structure cannot, in general, be realized in a purely quadratic GKLS model by simply retuning bath occupations, illustrating the genuinely interacting nature of the resetting dynamics.
  }
  \label{fig:two-site-nss}
\end{figure}
In the simulations we use parameters \( \varepsilon_1=0,\quad\varepsilon_2=0.5,\quad J=0.3,\quad \gamma_1=\gamma_2=0.5,\quad
f_1=0.2,\quad f_2=0.8,\) and monitor the time evolution for different interaction strengths $U$. The initial SPDM is chosen as
\begin{equation}
V(0) =
\begin{pmatrix}
0 & 0.1\\
0.1 & 1
\end{pmatrix},
\end{equation}
which is then projected onto the physical region by eigenvalue clipping.

Figure~\ref{fig:two-site-n1t} shows $n_1(t)$ for four different interaction strengths $U=0,0.5,1.0,2.0$. For $U=0$ the system relaxes smoothly to a steady state determined entirely by the noninteracting GKLS Liouvillian. As $U$ increases, the transient dynamics develops interaction-dependent dephasing and the asymptotic occupation $n_1^{\mathrm{ss}}(U)$ changes. The interaction thus modifies both the approach to steady state and the steady-state value itself.

To extract the steady state more quantitatively, we integrate the equations to a long time $T$ for a range of $U$ values and identify the approximate steady-state SPDM $V^*(U)=V(T)$. Figure~\ref{fig:two-site-nss} shows the resulting steady-state occupations $n_1^{\mathrm{ss}}(U)=V^*_{11}(U)$ and $n_2^{\mathrm{ss}}(U)=V^*_{22}(U)$ as functions of $U$. Both occupations depend nonlinearly on $U$ and move in opposite directions. The pair $(n_1^{\mathrm{ss}}(U),n_2^{\mathrm{ss}}(U))$ traces out a curve in the $(n_1,n_2)$ plane which is a family of Hartree fixed points.

Importantly, these Hartree fixed-point curves cannot, in general, be obtained from any noninteracting GKLS model with simply retuned bath occupations. In a noninteracting model, the steady state is the unique solution of a linear Lyapunov equation determined by $M$, $\Gamma$, and $f$. Changing $f$ moves the steady state in a restricted way. Here the Hartree feedback modifies the effective on-site energies in a state-dependent manner, leading to a nonlinear fixed-point equation for the occupations. The interaction thus enables steady-state structures that are absent in the purely quadratic resetting model.

\begin{figure}[t]
    \centering
    \subfigure[]{\includegraphics[width=0.235\textwidth]{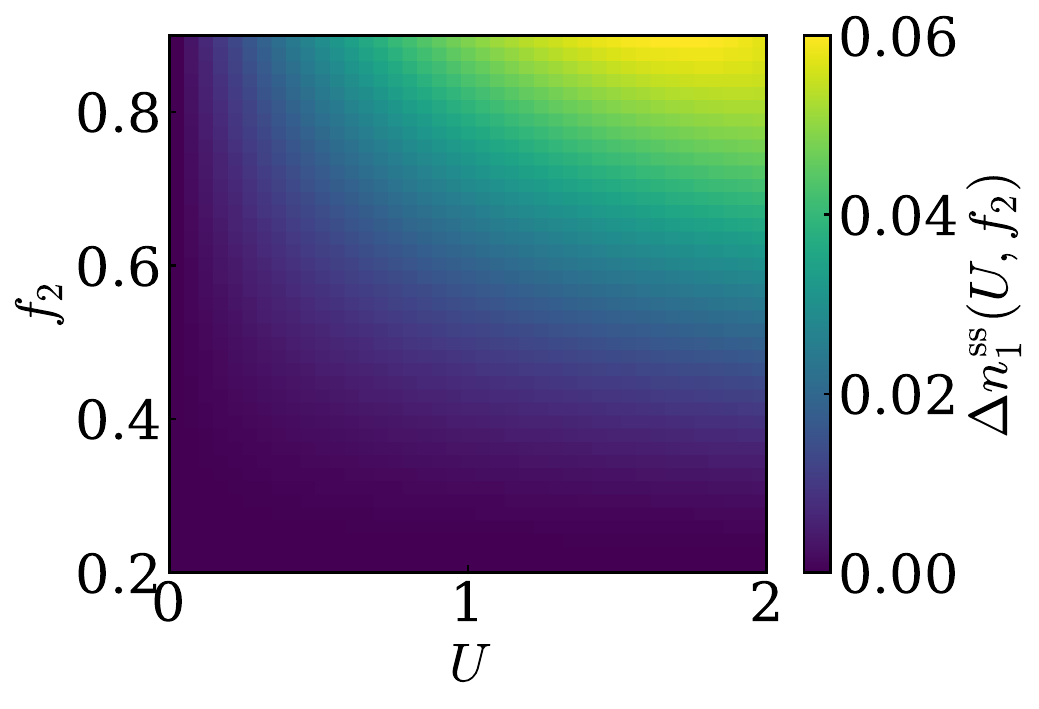}} 
    \subfigure[]{\includegraphics[width=0.235\textwidth]{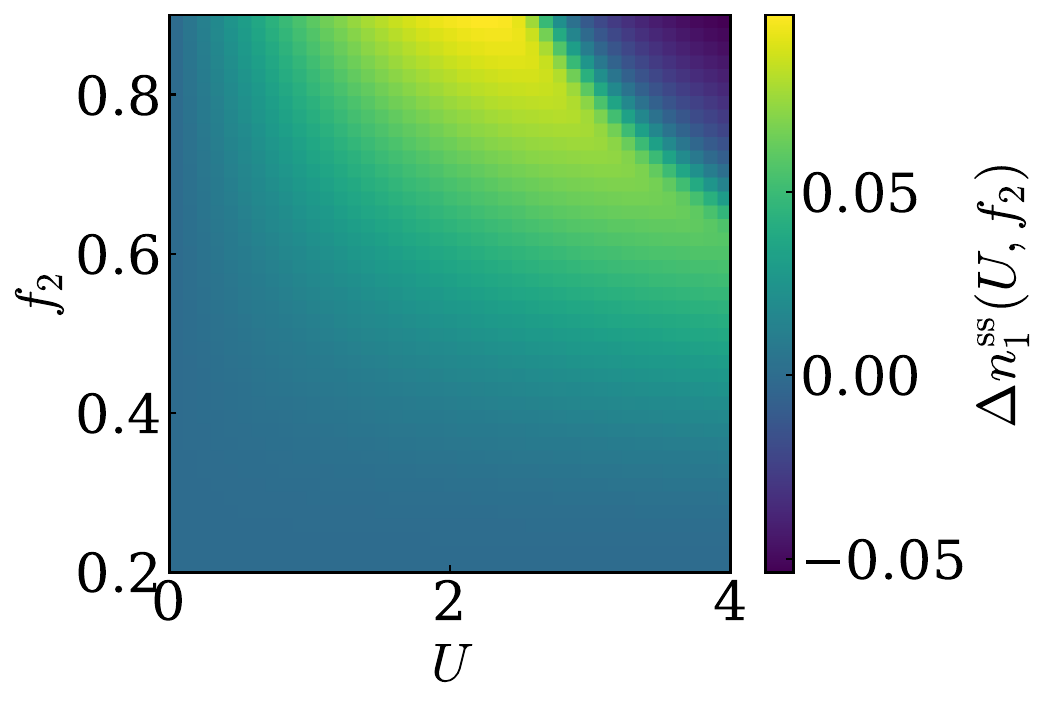}} 
  \caption{
    Interaction-induced change of the steady-state occupation on the left site in the two-site GKLS+Hartree model.
    We plot $\Delta n_1^{\mathrm{ss}}(U,f_2)
    = n_1^{\mathrm{ss}}(U,f_2) - n_1^{\mathrm{ss}}(0,f_2)$ as a function of the interaction strength $U$ and the right-bath occupation $f_2$, with the left bath fixed at $f_1=0.2$.
    (a) Default parameter regime with $\varepsilon_1 = 0$, $\varepsilon_2 = 0.5$, and $J = 0.3$.
    Here $\Delta n_1^{\mathrm{ss}}$ is non-negative and grows smoothly with both $U$ and $f_2$, indicating a bias-dependent enhancement of the left-site occupation due to interactions.
    (b) Resonant regime with $\varepsilon_1 = -0.4$, $\varepsilon_2 = +0.4$, and $J = 0.25$.
    In this case $\Delta n_1^{\mathrm{ss}}(U,f_2)$ develops a ridge-and-valley structure and changes sign: for moderate $U$ and large $f_2$ interactions strongly enhance $n_1^{\mathrm{ss}}$, whereas for larger $U$ they suppress it below the quadratic value.
    This non-monotonic, sign-changing landscape cannot be generated by simply retuning bath parameters in a purely quadratic GKLS model and is a clear signature of interaction-enabled Hartree fixed points.
  }
  \label{fig:delta-n1ss-U-f2}
\end{figure}

To visualize the interaction-enabled steady-state structure more globally, in Fig.~\ref{fig:delta-n1ss-U-f2} we plot the interaction-induced change
$\Delta n_1^{\mathrm{ss}}(U,f_2)
 = n_1^{\mathrm{ss}}(U,f_2) - n_1^{\mathrm{ss}}(0,f_2)$
as a function of $U$ and the right-bath occupation $f_2$.
In the default parameter regime [Fig.~\ref{fig:delta-n1ss-U-f2}(a)] $\Delta n_1^{\mathrm{ss}}$ is non-negative and increases smoothly with both $U$ and $f_2$, showing that interactions act as a bias-dependent renormalization of the steady-state response.
In the resonant regime [Fig.~\ref{fig:delta-n1ss-U-f2}(b)], however, $\Delta n_1^{\mathrm{ss}}(U,f_2)$ develops a pronounced ridge-and-valley structure and even changes sign, with interactions first enhancing and then suppressing $n_1^{\mathrm{ss}}$ as $U$ is increased. Such a sign-changing, non-monotonic dependence on $(U,f_2)$ cannot be reproduced by any quadratic quasi-free GKLS model with retuned bath parameters, and directly reflects the interaction-enabled Hartree fixed points of the two-site system.

\section{Summary and Discussion}
\label{sec:discussion}

We have developed a controlled framework for incorporating weak interactions into fermionic resetting dynamics while preserving complete positivity and quasi-free structure at the SPDM level. Starting from a quadratic ring and a stroboscopic resetting protocol, we derived the subsystem SPDM map and its continuous-time affine structure. By introducing density-density interactions and treating them at Hartree level, we obtained a nonlinear, but closed, SPDM equation with a state-dependent effective Hamiltonian.
The key technical step was a constructive CP-safe Gaussian embedding theorem showing that any affine SPDM evolution of the form~\eqref{eq:SPDM-affine-main} arises from a quadratic fermionic GKLS master equation with local jump operators.
This embedding naturally extends to the Hartree case when the effective single-particle Hamiltonian is used as a time-dependent Hermitian Hamiltonian in the Lindblad generator.

The Hartree approximation preserves the structural closure of the quadratic resetting model while introducing genuinely nonlinear dependencies on local occupations. The CP-safe embedding shows that resetting and GKLS dissipation provide consistent
continuous- and discrete-time descriptions of the same physical three-layer architecture: the subsystem $S$, a finite environment $E$, and an external thermal reservoir acting only on $E$.  The resulting dynamics exhibits interaction-enabled steady states and transient structures that cannot arise in purely quadratic models, demonstrating that weak interactions qualitatively expand the phenomenology accessible to resetting-based approaches.  This framework provides a robust foundation for studying interacting nonequilibrium states, dynamical fixed points, and transport in extended geometries.

Our numerical examples demonstrate that interactions qualitatively modify the nonequilibrium steady states of resetting dynamics.
In the ring geometry, the steady-state plateau of the subsystem occupation becomes interaction-dependent and is accurately captured by the GKLS embedding. In the minimal two-site model, interactions generate a family of nonlinear Hartree fixed points for the SPDM that cannot be obtained in any noninteracting quasi-free model with merely retuned bath parameters. These examples highlight interaction-enabled physics beyond the quadratic resetting paradigm.

The present work opens several directions for future research. On the interaction side, one can go beyond simple on-site Hartree terms and incorporate Fock contributions or longer-range interactions, while still retaining closure at the SPDM level or at the level of higher-order covariance matrices. On the dissipative side, more general quasi-free Lindblad operators, including nonlocal jump operators and structured spectral densities, can be considered. An interesting avenue is to explore regimes where the Hartree nonlinearity leads to multiple steady states, dynamical bistability, or discontinuous transitions in the Liouvillian spectrum.
Such phenomena would provide sharp signatures of interaction-enabled nonequilibrium phases in resetting dynamics.

A related question concerns the role of correlations between the subsystem and the environment in resetting-type descriptions.  In the noninteracting quadratic resetting framework \cite{vieira2020}, the authors introduced two limiting protocols: the “repeated-interaction” (RI) scheme, in which both the environment block $V_{EE}$ and the $S–E$ coherences are discarded after each stroke, and the “evolving-correlations” (EC) scheme, in which only $V_{EE}$ is reset to the target $F_E$ while $V_{SE}$ and $V_{ES}$ are allowed to accumulate between strokes.  Our stroboscopic Hartree protocol on the ring is of RI type, since after each unitary step we reset $V_{EE}\to F_E$ and set $V_{SE}=V_{ES}=0$.  By contrast, the CP-safe Gaussian Lindblad embedding~\eqref{eq:SPDM-Hartree-CP-main} is closer in spirit to an EC protocol: only the environment sites are directly damped towards $F_E$, whereas $S–E$ coherences are generated coherently and decay on the dissipative time scale rather than being instantaneously erased.  In the quadratic limit the continuous-time generator obtained as the small–$\tau$ limit of an EC map belongs to the same affine SPDM family as Eq.~\eqref{eq:SPDM-affine-main}, so our GKLS construction can be viewed as a correlation-preserving, completely positive realisation of an EC-type resetting dynamics extended to include Hartree interactions (see Appendix ~\ref{app:EC-reset}).

Finally, the CP-safe Hartree embedding developed here is directly applicable to mesoscopic transport setups where a small interacting region is repeatedly coupled to finite reservoirs that are periodically reinitialized.
Our framework provides a bridge between repeated-interaction descriptions and Lindblad master equations, allowing one to study interaction effects in a transparent and computationally tractable way.

\section*{Data availability}

All data shown in the figures are available from the authors upon reasonable request.

\section{Acknowledgements}

J.K. and T.A-N. have been supported by the Academy of Finland through its QTF Center of Excellence grant no. 312298 and
by the European Union and the European Innovation Council through the Horizon Europe project QRC-4-ESP (grant No. 101129663), and EU Horizon Europe Quest project (No. 10116088). A.L. acknowledges support from the Leverhulme Trust Research Project Grant RPG-2025-063. 

\appendix

\section{Hamiltonian evolution of the SPDM}
\label{app:Hamiltonian-evolution}

In this appendix we derive Eq.~\eqref{eq:SPDM-unitary} in detail.
We consider the quadratic Hamiltonian
\begin{equation}
\hat H_0=\sum_{\mu\nu}a_\mu^\dagger M_{\mu\nu}a_\nu
\end{equation}
and the operator $O_{\alpha\beta}=a_\alpha^\dagger a_\beta$.
The Heisenberg equation for $O_{\alpha\beta}$ is
\begin{equation}
\frac{d}{dt}O_{\alpha\beta}=\frac{i}{\hbar}[\hat H_0,O_{\alpha\beta}].
\end{equation}
Using linearity of the commutator,
\begin{equation}
[\hat H_0,O_{\alpha\beta}]
=\sum_{\mu\nu}M_{\mu\nu}[a_\mu^\dagger a_\nu,a_\alpha^\dagger a_\beta].
\label{eq:A1}
\end{equation}
The basic commutator between fermionic bilinears can be computed using the canonical anticommutation relations.
We start from
\begin{equation}
[a_\mu^\dagger a_\nu,a_\alpha^\dagger a_\beta]
=a_\mu^\dagger a_\nu a_\alpha^\dagger a_\beta
-a_\alpha^\dagger a_\beta a_\mu^\dagger a_\nu.
\end{equation}
Using
\(
a_\nu a_\alpha^\dagger=\delta_{\nu\alpha}-a_\alpha^\dagger a_\nu
\)
and
\(
a_\beta a_\mu^\dagger=\delta_{\beta\mu}-a_\mu^\dagger a_\beta
\),
we find
\begin{align}
a_\mu^\dagger a_\nu a_\alpha^\dagger a_\beta
&= a_\mu^\dagger(\delta_{\nu\alpha}-a_\alpha^\dagger a_\nu)a_\beta
\nonumber\\
&= \delta_{\nu\alpha}a_\mu^\dagger a_\beta - a_\mu^\dagger a_\alpha^\dagger a_\nu a_\beta,
\\
a_\alpha^\dagger a_\beta a_\mu^\dagger a_\nu
&= a_\alpha^\dagger(\delta_{\beta\mu}-a_\mu^\dagger a_\beta)a_\nu
\nonumber\\
&= \delta_{\beta\mu}a_\alpha^\dagger a_\nu - a_\alpha^\dagger a_\mu^\dagger a_\beta a_\nu.
\end{align}
Subtracting these expressions gives
\begin{align}
[a_\mu^\dagger a_\nu,a_\alpha^\dagger a_\beta]
&= \delta_{\nu\alpha}a_\mu^\dagger a_\beta
-\delta_{\beta\mu}a_\alpha^\dagger a_\nu
\nonumber\\
&\quad -a_\mu^\dagger a_\alpha^\dagger a_\nu a_\beta
+a_\alpha^\dagger a_\mu^\dagger a_\beta a_\nu.
\label{eq:A2}
\end{align}
The last two terms cancel because swapping $a_\mu^\dagger$ and $a_\alpha^\dagger$ produces a minus sign, and likewise for $a_\nu$ and $a_\beta$.
Explicitly,
\begin{equation}
a_\mu^\dagger a_\alpha^\dagger a_\nu a_\beta
=-a_\alpha^\dagger a_\mu^\dagger a_\nu a_\beta
=+a_\alpha^\dagger a_\mu^\dagger a_\beta a_\nu,
\end{equation}
so the difference in Eq.~\eqref{eq:A2} vanishes.
Therefore
\begin{equation}
[a_\mu^\dagger a_\nu,a_\alpha^\dagger a_\beta]
=\delta_{\nu\alpha}a_\mu^\dagger a_\beta
-\delta_{\beta\mu}a_\alpha^\dagger a_\nu.
\end{equation}
Inserting this back into Eq.~\eqref{eq:A1} yields
\begin{align}
[\hat H_0,O_{\alpha\beta}]
&= \sum_{\mu\nu}M_{\mu\nu}
\left(\delta_{\nu\alpha}a_\mu^\dagger a_\beta
-\delta_{\beta\mu}a_\alpha^\dagger a_\nu\right)
\nonumber\\
&= \sum_\mu M_{\mu\alpha}a_\mu^\dagger a_\beta
-\sum_\nu M_{\beta\nu}a_\alpha^\dagger a_\nu.
\end{align}
Taking expectation values with respect to $\rho(t)$ we obtain
\begin{equation}
i\hbar\frac{d}{dt}V_{\alpha\beta}
= \sum_\mu M_{\mu\alpha} V_{\mu\beta}
 -\sum_\nu M_{\beta\nu} V_{\alpha\nu},
\end{equation}
which in matrix notation is Eq.~\eqref{eq:SPDM-unitary}.

\section{Hartree contribution: detailed derivation}
\label{app:Hartree-details}

Here we provide additional intermediate steps for the Hartree mean-field derivation leading to Eq.~\eqref{eq:Hartree-component}.
We start from the interaction Hamiltonian
\[
\hat H_{\mathrm{int}}=\frac12\sum_{\gamma\delta}U_{\gamma\delta}\hat n_\gamma\hat n_\delta,
\]
with $\hat n_\gamma=a_\gamma^\dagger a_\gamma$.
The commutator with $O_{\alpha\beta}=a_\alpha^\dagger a_\beta$ is
\[
[\hat H_{\mathrm{int}},O_{\alpha\beta}]
=\frac12\sum_{\gamma\delta}U_{\gamma\delta}[\hat n_\gamma\hat n_\delta,O_{\alpha\beta}].
\]
Using $[AB,O]=A[B,O]+[A,O]B$ we find
\[
[\hat n_\gamma\hat n_\delta,O_{\alpha\beta}]
=\hat n_\gamma[\hat n_\delta,O_{\alpha\beta}]+[\hat n_\gamma,O_{\alpha\beta}]\hat n_\delta.
\]
The commutator of a number operator with $O_{\alpha\beta}$ follows from
\[
[\hat n_\gamma,a_\alpha^\dagger]=\delta_{\gamma\alpha}a_\alpha^\dagger,\qquad
[\hat n_\gamma,a_\beta]=-\,\delta_{\gamma\beta}a_\beta,
\]
which together imply
\begin{align}
[\hat n_\gamma,O_{\alpha\beta}]
&= [\hat n_\gamma,a_\alpha^\dagger]a_\beta + a_\alpha^\dagger[\hat n_\gamma,a_\beta]
\nonumber\\
&= \delta_{\gamma\alpha}a_\alpha^\dagger a_\beta - \delta_{\gamma\beta}a_\alpha^\dagger a_\beta
\nonumber\\
&= (\delta_{\gamma\alpha}-\delta_{\gamma\beta})O_{\alpha\beta}.
\end{align}
Therefore
\begin{align}
[\hat n_\gamma\hat n_\delta,O_{\alpha\beta}]
&= (\delta_{\delta\alpha}-\delta_{\delta\beta})\hat n_\gamma O_{\alpha\beta}\notag\\
&+(\delta_{\gamma\alpha}-\delta_{\gamma\beta})O_{\alpha\beta}\hat n_\delta.
\end{align}
Inserting this into the commutator with $\hat H_{\mathrm{int}}$ gives Eq.~\eqref{eq:Hartree-component} in the main text.
The Hartree factorization and subsequent steps are then straightforward.

\section{Dissipative SPDM evolution: single-mode and many-mode}
\label{app:dissipator-details}

In this appendix we derive Eq.~\eqref{eq:Vdot-diss-main}.
We first consider a single fermionic mode and then generalize to many modes.

\subsection{Single-mode case}

For a single mode we drop the site index and consider jump operators
\(
L^-=\sqrt{\gamma(1-f)}a
\)
and
\(
L^+=\sqrt{\gamma f}a^\dagger
\).
The number operator is $\hat n=a^\dagger a$ and the expectation value $n(t)=\Tr[\rho(t)\hat n]$.
The dissipative contribution to $\dot n$ from $L^-$ is
\begin{align}
\dot n\Big|_{L^-}
&= \Tr\!\left(\mathcal D[L^-](\rho)\hat n\right)
\nonumber\\
&= \gamma(1-f)\left(\Tr(a\rho a^\dagger\hat n)
-\frac12\Tr(\{a^\dagger a,\rho\}\hat n)\right).
\end{align}
Using cyclicity of the trace, the first term becomes
\[
\Tr(a\rho a^\dagger\hat n)=\Tr(\rho a^\dagger\hat n a).
\]
Since $\hat n=a^\dagger a$, we have $a^\dagger\hat n a=a^\dagger a^\dagger a a=0$ because $a^\dagger a^\dagger=0$ for fermions.
Thus $\Tr(a\rho a^\dagger\hat n)=0$.
For the anticommutator term we use
\(
\{a^\dagger a,\rho\}=a^\dagger a\rho +\rho a^\dagger a
\),
so
\begin{align}
\Tr(\{a^\dagger a,\rho\}\hat n)
&= \Tr(a^\dagger a\rho\hat n)+\Tr(\rho a^\dagger a\hat n)
\nonumber\\
&= \Tr(\rho\hat n a^\dagger a)+\Tr(\rho a^\dagger a\hat n)
\nonumber\\
&= 2\Tr(\rho\hat n^2).
\end{align}
Since $\hat n^2=\hat n$ for a fermionic mode, this equals $2n$.
We thus obtain
\[
\dot n\Big|_{L^-}=-\gamma(1-f)n.
\]

For $L^+=\sqrt{\gamma f}a^\dagger$ the dissipative contribution is
\begin{align}
\dot n\Big|_{L^+}
&= \gamma f\left(\Tr(a^\dagger\rho a\hat n)
-\frac12\Tr(\{aa^\dagger,\rho\}\hat n)\right).
\end{align}
The first term can be written as
\[
\Tr(a^\dagger\rho a\hat n)=\Tr(\rho a\hat n a^\dagger).
\]
Using $\hat n=a^\dagger a$ and the anticommutator relations one shows that $a\hat n a^\dagger=1-\hat n$, so
\[
\Tr(a^\dagger\rho a\hat n)=\Tr(\rho(1-\hat n))=1-n.
\]
For the anticommutator term we use $aa^\dagger=1-\hat n$ and obtain
\begin{align}
\Tr(\{aa^\dagger,\rho\}\hat n)
&=\Tr((1-\hat n)\rho\hat n)+\Tr(\rho(1-\hat n)\hat n)
\nonumber\\
&=0,
\end{align}
because $(1-\hat n)\hat n=0$.
Thus
\[
\dot n\Big|_{L^+}=\gamma f(1-n).
\]
Adding the contributions from $L^-$ and $L^+$ gives
\[
\dot n=-\gamma(1-f)n + \gamma f(1-n) = -\gamma(n-f).
\]

\subsection{Many-mode case}

For many modes with local jump operators $L_\alpha^\pm$ one finds that each diagonal SPDM element $V_{\alpha\alpha}=\langle \hat n_\alpha\rangle$ obeys the same equation with rate $\gamma_\alpha$ and target $f_\alpha$:
\[
\dot V_{\alpha\alpha}=-\gamma_\alpha(V_{\alpha\alpha}-f_\alpha).
\]
Off-diagonal elements $V_{\alpha\beta}$ with $\alpha\neq\beta$ are damped by both sites and satisfy
\[
\dot V_{\alpha\beta}=-\frac12(\gamma_\alpha+\gamma_\beta)V_{\alpha\beta}.
\]
These equations are conveniently encoded in the matrix form of Eq.~\eqref{eq:Vdot-diss-main}.

\section{Two-site equations and numerical scheme}
\label{app:two-site-equations}

Here we summarize the two-site SPDM equations in component form and the numerical algorithm used to generate Figs.~\ref{fig:two-site-n1t} and~\ref{fig:two-site-nss}.
Writing
\[
V=\begin{pmatrix}
n_1 & c\\
c^* & n_2
\end{pmatrix},
\]
and
\[
M_{\mathrm{eff}}(V)=
\begin{pmatrix}
\varepsilon_1+U n_2 & -J\\
-J & \varepsilon_2+U n_1
\end{pmatrix},
\]
we insert these into Eq.~\eqref{eq:SPDM-Hartree-CP-main}.
The Hamiltonian contribution is $(i/\hbar)[V,M_{\mathrm{eff}}(V)]$.
A straightforward calculation gives
\begin{align}
\dot n_1\Big|_H &= \frac{2J}{\hbar}\Im c,\\
\dot n_2\Big|_H &= -\frac{2J}{\hbar}\Im c,\\
\dot c\Big|_H &= \frac{i}{\hbar}\left[(\varepsilon_2+U n_1) - (\varepsilon_1+U n_2)\right]c + \frac{iJ}{\hbar}(n_2-n_1).
\end{align}
The dissipative part is $-\tfrac12\{\Gamma,V\}+\Gamma f$ with $\Gamma=\mathrm{diag}(\gamma_1,\gamma_2)$ and $f=\mathrm{diag}(f_1,f_2)$.
This yields
\begin{align}
\dot n_1\Big|_{\mathrm{diss}} &= -\gamma_1(n_1-f_1),\\
\dot n_2\Big|_{\mathrm{diss}} &= -\gamma_2(n_2-f_2),\\
\dot c\Big|_{\mathrm{diss}} &= -\frac12(\gamma_1+\gamma_2)c.
\end{align}
Adding the two contributions gives Eqs.~\eqref{eq:n1-eq}--\eqref{eq:c-eq} in the main text.

For the numerical integration we use an explicit Euler step of size $\Delta t$:
\begin{align}
n_1(t+\Delta t) &= n_1(t)+\Delta t\,\dot n_1(t),\\
n_2(t+\Delta t) &= n_2(t)+\Delta t\,\dot n_2(t),\\
c(t+\Delta t)   &= c(t)+\Delta t\,\dot c(t),
\end{align}
where $\dot n_1$, $\dot n_2$, and $\dot c$ are evaluated from Eqs.~\eqref{eq:n1-eq}--\eqref{eq:c-eq}.
After each step we reconstruct $V(t+\Delta t)$, symmetrize it as $V\to\tfrac12(V+V^\dagger)$, diagonalize it, clip its eigenvalues to the interval $[0,1]$, and reconstruct $V$.
This ensures that $V$ remains a valid SPDM with eigenvalues between $0$ and $1$.

\section{Ring numerics}
\label{app:ring-numerics}

For completeness we briefly summarize the numerical procedure used for the ring segmentation examples in Sec.~\ref{sec:numerics-main}.
We consider a ring with $N=N_S+N_E$ sites described by a tight-binding Hamiltonian with nearest-neighbor hopping $J$ and on-site energies $\varepsilon_\alpha$.
The subsystem $S$ consists of $N_S$ contiguous sites; the environment $E$ contains the remaining $N_E$ sites.
We include on-site Hartree interactions $U_{\gamma\delta}=U\delta_{\gamma\delta}$ on all sites.

For the resetting dynamics, each step consists of:
(i) replacing the environment block $V_E$ of the SPDM by the thermal target $F_E$, and setting system--environment correlations $C$ to zero; and
(ii) evolving the full SPDM unitarily for a time $\tau$ under the mean-field Hamiltonian with current Hartree potential.
The unitary evolution is implemented at the SPDM level using $V\mapsto U(V) V U^\dagger(V)$ with $U(V)=\exp[-(i/\hbar)M_{\mathrm{eff}}(V)\tau]$.
The average subsystem occupation $\langle n_S\rangle$ is recorded at each reset.

For the GKLS embedding we solve Eq.~\eqref{eq:SPDM-Hartree-CP-main} for the full $N\times N$ SPDM using an explicit Euler method with time step $\Delta t$.
The damping matrix $\Gamma$ is diagonal and nonzero only on environment sites; the target $f$ coincides with $F_E$ on $E$ and vanishes on $S$.
After each step we enforce Hermiticity and clip eigenvalues as in the two-site model.
The subsystem occupation $\langle n_S(t)\rangle$ is computed from the subsystem diagonal block of $V(t)$.

\section{EC-type resetting and affine SPDM generators}
\label{app:EC-reset}

For completeness we briefly recall the EC (“evolving-correlations”) variant of the quadratic resetting protocol and its relation to the affine SPDM generators used in the main text.  We split the single-particle density matrix into subsystem and environment blocks,
\begin{equation}
  V =
  \begin{pmatrix}
    V_{SS} & V_{SE} \\
    V_{ES} & V_{EE}
  \end{pmatrix},
\end{equation}
with $S$ the first $N_S$ sites and $E$ the remaining $N_E$ sites.  Starting from $V^{(n)}$ at stroboscopic time $t=n\tau$, one unitary stroke with a quadratic Hamiltonian $H_0$ produces
\begin{equation}
  \tilde V = U V^{(n)} U^\dagger, \qquad
  U = e^{-i H_0 \tau/\hbar}.
\end{equation}
In the RI protocol used in Sec.~\ref{sec:model} the reset step sets $V_{SE}=V_{ES}=0$ and $V_{EE}\to F_E$, leaving only $V_{SS}$ unchanged. In the EC protocol, by contrast, only the environment block is overwritten while subsystem–environment coherences are preserved:
\begin{align}
  V^{(n+1)}_{SS} = \tilde V_{SS}, \qquad
  V^{(n+1)}_{SE} = \tilde V_{SE}, \nonumber\\
  V^{(n+1)}_{ES} = \tilde V_{ES}, \qquad
  V^{(n+1)}_{EE} = F_E .
  \label{eq:EC-reset-rule}
\end{align}
For a quadratic $H_0$ and a diagonal target $F_E$, the resulting discrete map $V^{(n)} \mapsto V^{(n+1)}$ is quasi-free and can be written in the affine form $V \mapsto A V A^\dagger + B$.  Expanding for small $\tau$ one obtains a continuous-time SPDM equation of the type
\begin{equation}
  \dot V = \frac{i}{\hbar}[V,M] - \frac{1}{2}\{\Gamma,V\} + \Gamma f,
  \label{eq:EC-affine-generator}
\end{equation}
where $M$ is an effective single-particle Hamiltonian, $\Gamma\ge 0$ is supported on $E$, and $f$ encodes the target occupations on the environment sites.  Equation~\eqref{eq:EC-affine-generator} has precisely the affine SPDM structure realised by a quadratic Gaussian GKLS master equation, and is the quadratic ancestor of the CP-safe Hartree equation~\eqref{eq:SPDM-Hartree-CP-main} used in the main text.  In this sense our GKLS construction provides a continuous-time, completely positive embedding of an EC-type resetting dynamics in the presence of weak Hartree interactions.

\bibliography{references_hartree}     

\end{document}